# Communities of practice in the distributed international environment

*Paul Hildreth*
**Paul Hildreth** is a Research Student, Department of Computer Science, University of York, UK.

*Chris Kimble*

### The authors

**Paul Hildreth** is a Research Student, and **Chris Kimble** and **Peter Wright** are both Lecturers all at the Department of Computer Science, University of York, UK.

### Abstract

Modern commercial organisations are facing pressures which have caused them to lose personnel. When they lose people, they also lose their knowledge. Organisations also have to cope with the internationalisation of business forcing collaboration and knowledge sharing across time and distance. Knowledge management (KM) claims to tackle these issues. This paper looks at an area where KM does not offer sufficient support, that is, the sharing of knowledge that is not easy to articulate. The focus in this paper is on communities of practice in commercial organisations. We do this by exploring knowledge sharing in Lave and Wenger's (1991) theory of communities of practice and investigating how communities of practice may translate to a distributed international environment. The paper reports on two case studies that explore the functioning of communities of practice across international boundaries.

### Electronic access

The current issue and full text archive of this journal is available at
**http://www.emerald-library.com**

Chris Kimble is a Lecturer at the Department of Computer Science, University of York, UK.

*Peter Wright*
Peter Wright is a Lecturer at the Department of Computer Science, University of York, UK.

## Introduction

As globalisation affects business, many organisations have taken steps to downsize, outsource and deskill in an effort to remain competitive (Davenport and Prusak, 1998; O'Dell and Jackson Grayson, 1998). This has had an unexpected effect. Both downsizing and outsourcing mean a reduction in staffing levels. As people have left, companies have realised that with them they have taken a valuable stock of knowledge.

This knowledge is increasingly seen as central to the success of organisations and an asset that needs to be managed (Boersma and Stegwee, 1996). It can be both knowledge of how the work is done in practice, and knowledge of a particular domain (Sachs, 1995). Domain knowledge is relatively easy to replace. The knowledge of how a company operates built up over long periods of experience is in many cases irreplaceable at least in the short term.

In this paper, we will differentiate between "hard" and "soft" knowledge. Hard knowledge is knowledge that can be easily articulated and captured. Soft knowledge on the other hand is not so easily articulated and cannot be so readily captured. Examples of soft knowledge might be:
- experience;
- work knowledge which has been internalised;
- tacit knowledge (Nonaka, 1991) e.g. using a word processor (Goguen, 1997).

The management of hard knowledge is well established but the sharing of soft knowledge poses greater problems. One of the important issues for knowledge management (KM) is how can this soft knowledge be shared when it is so difficult to articulate? One approach that may be useful can be found in the concepts of communities of practice (CoPs) and pegitimate peripheral participation (LPP) (Lave and Wenger, 1991).

## CoPs and legitimate peripheral participation

Lave and Wenger (1991) first introduced the concept of a CoPs in 1991. Although often seen as a simple apprenticeship model where soft knowledge is transferred through the situated learning that takes place in apprenticeships, the central concept of legitimate peripheral participation was not restricted to apprenticeships alone.

Lave and Wenger (1991) described a CoP as "…a set of relations among persons, activity, and world, over time and in relation with other tangential and overlapping CoPs". In these communities, newcomers learn from old-timers by being allowed to participate in certain tasks relating to the practice of the community. Over time newcomers move from peripheral to full participation in the community.

Lave and Wenger (1991) saw a CoP as "an intrinsic condition for the existence of knowledge". They saw the learning that took place in such communities not as narrow situated learning where instances of practice are simply replicated but "learning as legitimate peripheral participation". LPP is not merely learning situated in practice but learning as an integral part of practice: learning as "generative social practice in the lived in world".

LPP is complex and composite in character. They state that each of its three aspects legitimation, peripherality and participation are indispensable in defining the others and can not be considered in isolation. Legitimation and participation define the characteristic ways of belonging to a community while peripherality and participation are concerned with location and identity in the social world.

Although Lave and Wenger (1991) stress the composite character of LPP, it is useful as an analytical convenience to consider the three components and their relationships separately.

Legitimation is the dimension of CoPs that is concerned with power and authority relations in the group. In the studies (non-drinking alcoholics, Goa tailors, quartermasters, butchers and Yucatan midwives), legitimation does not necessarily have to be formal. For example for quartermasters, tailors and butchers there is a degree of formal legitimacy that comes from hierarchy and rank but for the midwives and alcoholics legitimacy is more informal. For example, the alcoholics gain legitimacy, as the stories they tell of their experiences become more mature and closer to those of an old-timer.

Peripherality is not a physical concept as in core and periphery nor a simple measure of the amount of knowledge that has been acquired. Lave and Wenger (1991) use the terms peripheral and full participation to denote the degree of engagement with and participation in the community but note that peripherality "must be connected to issues of legitimacy of the social organisation and control over resources if it is to gain its full analytical potential".

For Lave and Wenger (1991), participation provides the key to understanding CoPs. CoPs do not necessarily imply co-presence, a well-defined or identifiable group, or socially visible boundaries. However, CoPs do imply participation in an activity about which all participants have a common understanding about what it is and what it means for their lives and community. The community and the degree of participation in it are in some senses inseparable from the practice.

**Extending the CoP concept**
In order to work effectively in a distributed international environment, companies are increasingly turning to international teams (Castells, 1996; Lipnack and Stamps, 1997; West *et al.*, 1997). These are seen as an effective and flexible means of bringing both skills and expertise to specific problems and tasks (Lotus, a and b). In response to this, the notion of CoPs has been extended from Lave and Wenger's (1991) model to encompass a wider range of definitions (Manville and Foote, 1996; Stewart, 1996; Orr, 1990; Seely Brown and Duguid, 1996).

There have been several attempts to define CoPs in the commercial environment and even some attempts by consultancies to formalise them. Manville and Foote (1996) offered the following definition

> ...a group of professionals informally bound to one another through exposure to a common class of problems, common pursuit of solutions, and thereby themselves embodying a store of knowledge (Manville and Foote, 1996).

Seely Brown and Solomon Gray (1998) took this further:

> At the simplest level, they are a small group of people who've worked together over a period of time. Not a team, not a task force, not necessarily an authorised or identified group They are peers in the execution of "real work". What holds them together is a common sense of purposes and a real need to know what each other knows.

Seely Brown and Duguid (1991) applied Lave and Wenger's (1991) ideas to an ethnographical study previously undertaken by Orr (1990). In this work, Orr studied a group of copier repair engineers from the perspective of their collective memory. Orr's explanation of how repairers fixed machines was based on the technicians' ability to share tacit knowledge.

Orr (1990) described how a technician could not complete a particularly difficult repair by simply following the manual. The technician called his supervisor and the two worked together until they had solved the problem. They did this by telling each other about similar problems they had encountered. The story-telling process enabled them to exchange their tacit knowledge and arrive at a solution to the problem. Over time, this solution was passed around the technicians and became part of the stock of knowledge of the community. Not only had they solved the problem but they had also contributed new knowledge and contributed to the development of the community.

An interesting aspect of this study was the use of narration. This has been commented on elsewhere (Seely Brown and Solomon Gray, 1998; Goldstein, 1993). Stories can be used to show the transition from newcomer to old-timer. As a newcomer's stories become accepted and develop into those of an established member he/she becomes a legitimate member of the community as in the non-drinking alcoholics of Lave and Wenger (1991). This in turn shapes the knowledge of the group, for it is the stories of those established members which engender the most confidence and which are accorded the greatest legitimacy.

**Are CoPs the same as teams?**
As we have seen, the term CoP has been extended to encompass new meanings that were not part of Lave and Wenger's (1991) original idea. This has led to the term "CoP" being applied, sometimes erroneously, to a wide range of groups, from project teams (Lindstaedt, 1996) to functional departments (Sandusky, 1997).

In this paper, we will use a distinction in the form of legitimation that is present to differentiate between a team and a CoP. In a team, we see legitimation as being derived from the formal hierarchy (e.g. externally imposed structure and membership). In CoPs legitimation is more informal and comes about by members earning their status in the community. This might be by the newcomer being accepted and gradually working his/her way to full participation.

It is possible for a team to become a CoP as informal relationships begin to develop and the source of legitimation changes in emphasis. Hutchins (1990, 1995) provides an account of how a formally structured team may also function as a CoP in his study of a navigation team on an American warship. There is a formal structure to the team provided by military rankings. However, when the team gets a new officer the informal CoP provides the forum for the learning that takes place. It is one of the petty officers, lower in rank but with more experience, who has to supervise the newcomer and "break in" a new officer.

**Can a CoP be virtual?**
The focus in this paper is on CoPs in commercial organisations. As we have seen, many commercial organisations now operate in a distributed, international environment. Hence, in order for such communities to function, they will have to operate (at least in part) in the virtual world.

Lave and Wenger's (1991) and Seely Brown and Duguid's (1991) examples of CoPs are co-located. However, the internationalisation of business, which is making companies turn to teams and communities, is also making operations more geographically distributed. This raises the question as to whether CoPs can continue to operate in such an environment, i.e. can a CoP be virtual?

There has been much discussion of virtual communities where the members never meet (Castells, 1996; Fernback, 1997; Poltrock and Engelbeck, 1997). Conkar et al. (1999) have discussed a multi-user dungeon (MUD) and referred to their members as a CoP. Although MUDs may appear to be an example of wholly virtual CoPs in fact, they are more similar to Lave and Wenger's (1991) CoPs. In a MUD, the medium is once again the practice. The MUD is not simply the medium by which the community communicates but it is also the reason for the existence of the community.

Computer supported co-operative work (CSCW) has also explored electronic support for distributed groups and teams. This work has resulted in a plethora of different terms such as virtual teams (Lipnack and Stamps, 1997), self-managing work groups (Williams, 1994; Evans and Sims, 1997) and Network Communities (Carroll et al., 1996).

Some aspects of a CoP should translate from the co-located to the virtual world relatively easily, for example, finding a common purpose or at least a shared interest. If the members are doing similar jobs then there will already be a shared domain language and knowledge.

We have seen that narration and the telling of stories can be used for sharing soft knowledge. At first glance, it should be easy to transfer stories to a distributed operation by simply recording the stories and making them available to members. However, this may not be so simple. The stories are not simply soft knowledge made hard. The listener also needs his/her soft knowledge in order to interpret the stories either to understand them or to make new inferences from them. This would also demonstrate a difference between a newcomer and an old-timer – the newcomer would just have the domain knowledge to understand some of the story. An old-timer would have the ability to make new interpretations.

Other issues might concern the question of how Lave and Wenger's (1991) concept of LPP would translate to a distributed environment. The learning undertaken with LPP is situated, as is some of the knowledge created during problem solving. Whether the CoP moves easily to working in distributed mode might depend on the reason for the situatedness. If the members need to be co-located because they share resources such as a document, then it should translate to the distributed environment relatively easily. If however the learning is situated because the face-to-face element is essential for seeing and learning how the job is done, then the distribution will have more impact.

The concept of peripherality may also be affected. In Lave and Wenger's (1991) CoPs, the periphery is a social periphery. However, in a distributed environment, there will also be a physical and a temporal periphery which will also have certain connotations for the notion of participation.

The transition to a virtual environment also raises the question of whether it will be more difficult to gain legitimacy in such a community, but perhaps the most difficult area will be the facilitation of participation. Participation is central to the evolution of the community and to the creation of relationships that help develop the sense of trust and identity, that defines the community.

**Distributed cognition, boundary objects and geographically distributed CoPs**

There is a body of literature that may help us understand how CoPs share soft knowledge in a distributed environment. The field of distributed cognition is concerned with representations of knowledge in teams.

Distributed cognition focuses on a representation of knowledge and its implementation. Hutchins (1990, 1995) uses the example of team navigation on a modern naval vessel as being work distributed across a team. He provides examples of computational artefacts, for example an alidade and a nautical slide rule, which, in his view embody the knowledge of previous generations. These artefacts are used as representations of knowledge which allow it to be moved around. The navigation team as a whole however, works like a CoP. Individuals move across roles at different times and newcomers join the team as others leave.

Using distributed cognition gives us a different view on a CoP. We now have two useful and complementary views of a CoP. LPP is concerned with the social structure of the community and how newcomers learn, whereas distributed cognition is concerned more with the process of how the work gets done. Together these two ideas give us a functional view of the CoP as well as a social view.

Hutchins (1990, 1995) however focuses on externalised representations of knowledge. It does not really address the case of how an artefact might be used by different communities and may be interpreted differently by them. Star (1989) and Star and Griesemer (1989), on the other hand, are concerned with the distribution of artefacts across communities. Boundary objects are artefacts used by communities: they cross the boundaries between communities and retain their structure, but are interpreted differently by them. The notion of boundary objects was developed by Star (1989) and Star and Griesemer (1989) as a way to explain co-ordination work between communities. Sandusky (1997) later applied the idea to CoPs. Boundary objects demonstrate that, although knowledge may be embedded in artefacts, it is not a simple matter of capturing the knowledge and passing it on. There is still some abstraction and some of the soft knowledge gets lost in the process. Domain knowledge is needed to both understand and use the artefact.

Again, the notions of distributed cognition and boundary objects are complementary. distributed cognition concentrates on the "absolute meaning" of artefacts and representations, whereas boundary objects are concerned with "interpretive flexibility" of representations across boundaries.

We will now use the ideas and concepts outlined above to interpret the results from two case studies that explore the functioning of CoPs that cross international boundaries.

**Report from real life**

We have seen the importance of CoPs to knowledge management in that they are groups where the sharing of soft knowledge takes place. It is also clear that companies have to operate in a distributed international environment and there may be problems with the transition from co-located communities to virtual communities. We have also seen that the concepts of distributed cognition and boundary objects may be of assistance in helping CoPs move into distributed international working. The next section reports on two case studies that explore distributed CoPs in an international environment. The first one looks to see if distributed international CoPs exist and the second explores the interactions and looks for shared artefacts.

**Case study one**

The question driving the first case study was could a CoP exist in the distributed international environment.

The case study was undertaken with Watson Wyatt, an international actuarial organisation and has been reported in detail elsewhere (Kimble *et al.*, 1998). It consisted of two parts. The first part was a questionnaire survey issued to all UK and European staff that aimed to collect factual information. This was followed by interviews to explore issues raised by the questionnaire.

Five metrics were selected as being indicators of CoPs but which were also testable in questionnaire format. These looked for people who:

**Figure 1** Evolution of the distributed community of practice

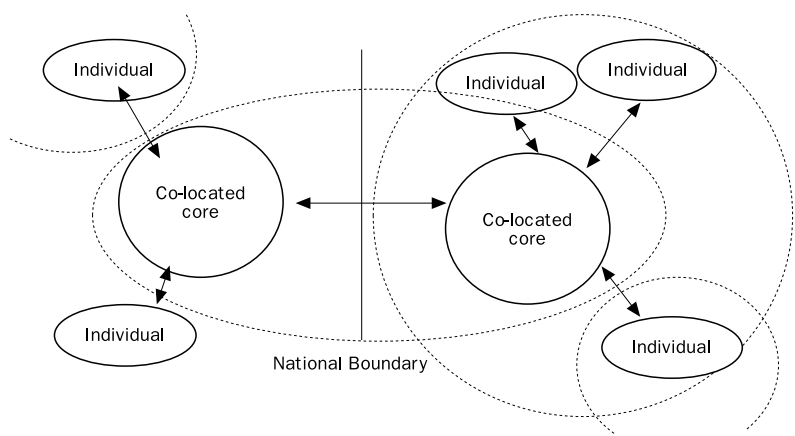

- were in regular contact with colleagues/peers doing similar jobs;
- talk with colleagues to solve problems;
- share projects with other colleagues;
- swap anecdotes/experiences with colleagues;
- learn from discussions with colleagues.

They were also divided into "same location" and "other locations" in order to differentiate between co-located and distributed communities. Those respondents which matched all five characteristics and which took place with other locations were extracted in order to focus on distributed CoPs. The interviews pursued CoPs in more depth, being able to seek a wider variety of characteristics.

The main outcome of the first case study was the existence of CoPs in the company and that they had a distributed aspect. There were however two other key points: evolution and degree of distribution.

*Evolution*
The CoPs found in the first case study appeared to evolve and develop through a three-stage process:
(1) The distributed CoP can evolve from an initial informal contact between its members or from an official grouping which becomes a CoP because of the way the members interact and work together.
(2) The co-located CoP may develop links with individuals in other locations who are doing similar work. These people may be members of other CoPs.
(3) The developing CoP may also link up with a similar group, possibly abroad.

Figure 1 shows two co-located cores that have developed a link between them. It also shows that there are possible links to individuals. The dotted lines show possible examples of CoPs.

*Distribution*
The findings of this first case study were encouraging in that they showed the existence of distributed CoPs. It also showed that, at least in the case of Watson Wyatt, they are not entirely distributed as in the MUD of Conkar *et al.* (1999). There appears to be a physically co-located element. Although LPP was central to the CoPs of Lave and Wenger (1991) it does not seem to be a key aspect of the distributed elements in the communities seen in the first case study. Where it did occur it was in a physically co-located part. This need for a co-located element supports findings from elsewhere (Lipnack and Stamps, 1997; Seely Brown and Duguid, 1996; Castells, 1996).

The structure of the CoP found in this first case study shows that there is a physical as well as a social periphery. There were some individuals who were in a distant location but felt themselves to be full members of the community but because of the distance they also felt they were on a physical periphery. They therefore did not have access to *ad hoc* encounters with their colleagues which those members who worked together in the co-located core were able to benefit from.

The first case study had demonstrated the existence of distributed CoPs but it also indicated that to explore in more detail the sharing of soft knowledge it would be necessary to examine the interaction and communication in such a group. The next section will report on a case study that was designed to do this.

**Figure 2** Structure of the group in case study two

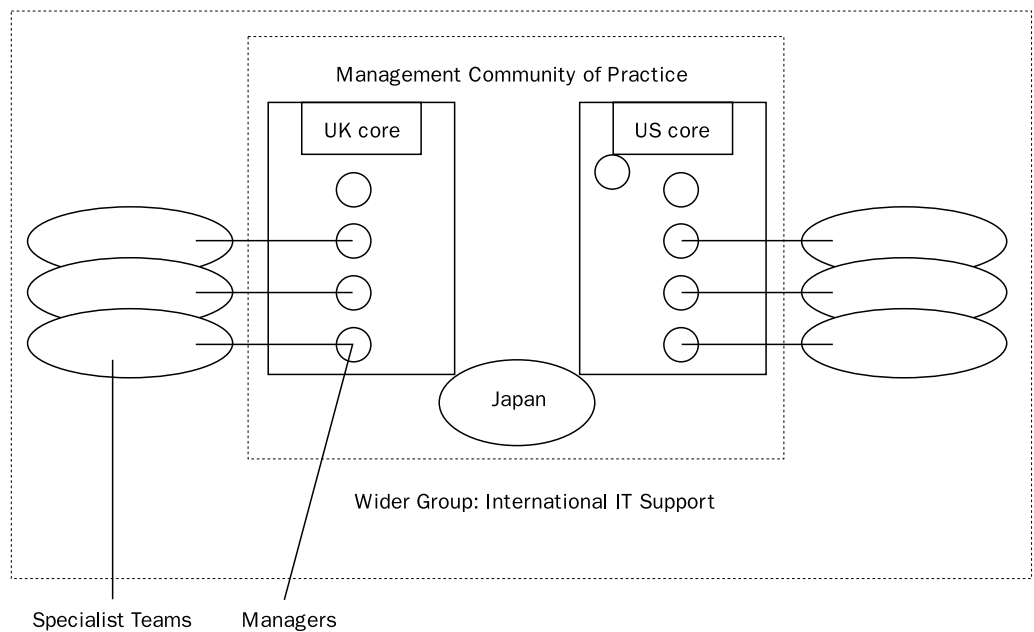

### Case study two

The second case study was undertaken with the research arm of a major international company. The group being investigated was the management team of IT support. The structure of the group matched the model created from the first case study. There was a group of four members co-located in the UK, a group of five members in the USA and one member in Japan (see Figure 2).

The UK team was identified as having a number of features that identified it as a CoP:

- a sense of common purpose;
- a strong feeling of identity;
- had its own terminology (group specific acronyms and nicknames);
- is an official group that evolved from a need but which is driven by the members themselves.

The evolution has continued developing the links with the US group to the point where the members consider themselves a CoP and have a feeling of identity to the point where they have given themselves a name. Within the group, there is a wide range of experience and knowledge for members to draw on. The culture of the company places great emphasis on working together and sharing knowledge. A focal point of the building was the café area where tea, coffee and biscuits were freely available and staff were encouraged to go there and sit and chat with their colleagues. Perhaps because of this a lot of the communication within the UK core is *ad hoc* and informal.

For exploring the interactions, communication, and collaboration, we must first understand it (Schmidt and Bannon, 1992). Suchman (1995) has also highlighted the importance of researching work in context. The case for ethnographical research for this kind of work has been recommended elsewhere (Cicourel, 1990). Case study two therefore used a participant observation approach to get first hand observation. A week was spent with the members of the UK core observing the day to day work and interactions of the members. The week spent with the group yielded several interesting insights into the workings of a distributed CoP.

*Relationships*

Of prime interest was how the group manages to function as a CoP in a distributed environment. Much of the work is undertaken within the cores but the members meet on a twice-yearly basis and, in between these meetings, maintain communication via e-media such as e-mail, voice mail, telephone, video link and Microsoft NetMeeting. The members of the group felt that during the face-to-face meetings they managed to get a lot of work done and develop much more quickly the relationships with their colleagues. During the periods of communication by e-media, they felt that the momentum gradually slowed until a physical meeting picked it up again.

The development of the relationships in a physical environment helped with issues of identity. The members felt they knew each other better more quickly than if they were developing the relationship via e-media. This then meant that they were confident with issues of identity during the time the work was being maintained through the e-media. Having a good personal relationship with the other members was seen as essential as it gave them confidence in what they were receiving from the other members, be it information, the solution to a problem or simply an opinion on an issue.

We were given two examples of relationships (outside of the community) which had developed over e-media. One of these had a serendipitous element about it and the other one had initially been adversarial until one of the communication partners happened to mention one of his interests. By chance this was a particular interest of his communication partner and fired a discussion. Over time, this grew into a strong relationship. However the feeling in the group was that these two examples were unusual and that a relationship can go further more quickly when it has face-to-face elements.

*Media*

The members of the group were generally very specific in their choice of media for certain tasks and as the face-to-face element was so important it was perhaps surprising that video conferencing was not the medium of choice for this distributed community as video is generally held to be the medium with the highest available bandwidth after face-to-face. The members' feeling was that the technology is still not ready and does not yet add sufficient extra, over and above a telephone link, to justify itself. Of far greater importance was speed of interaction. Consequently, telephone conferencing was a highly used medium often in conjunction with NetMeeting for sharing documents.

*Shared documents*

The sharing of documents proved to be a central activity during the week of the case study. We concentrated on one particular artefact, a planning document, that was being developed by the UK core of the group. This particular document was of interest because while it was being created for one prime purpose it was explicitly intended for some other purposes and also used for unintended purposes.

The planning document represented the application of the soft knowledge not only of the management team but also of their vertical teams. Each manager had input from his team in the form of an e-mail or a formatted document. They then merged this input and created a planning document of their own. At a management team meeting, these documents were discussed and one of the team then merged them into a draft planning document. During the week of the study, this draft document was the focus of three management team meetings during which it went through two more iterations. It was also used by at least one management team member to communicate with his team and drive a meeting.

The document was also the subject of informal *ad hoc* discussion and because of this, the structure was deliberately altered to take into account the need to communicate with the members in the USA. At the end of the week, there was a telephone conference with a one way video link and NetMeeting with some US members of the group. The planning document was the focus of this meeting and had been tailored for the purpose. During the meeting (and also during co-located meetings between UK members) by using the document, issues were raised for discussion and problems were flagged up – and solved, for example, as a result of discussing one of the points on the document a problem became apparent. The members of the group discussed the problem, applied their knowledge and experience and soon reached a solution. Similarly problems became apparent and one core found it was able to leverage from the experience of the other.

An important function of the document was its stimulative quality. It stimulated discussion of issues and the solving of problems (through either discussion or leverage) but also acted as a collaboration catalyst. The group used the document to highlight areas where they could collaborate on projects, where they could leverage from each other and where they could get their teams to work together. After the week of the case study, the planning document was merged with the US planning document to become one collaborative plan. The document in this latest form is still central to the work of the group and has become a living, ongoing document.

The document functioned as a boundary object with a difference. Boundary objects pass boundaries between communities (Sandusky, 1997) but it was interesting to note that the boundaries here as well as the physical boundaries between the cores were within the community. In this case, the boundaries were national and cultural boundaries and the document was specifically tailored for the purpose.

The findings in the second case study showed some interesting aspects of the work of the CoP in the distributed environment. These will be further considered in the next section.

## Discussion

**Relationships – confidence trust identity**

The importance of the face-to-face element even in a distributed CoP has some interesting implications. The strong personal relationship was felt to be essential to carry the community through the periods of e-communication. Knowing each other gave them a greater feeling of unity and common purpose or as one of the respondents put it, "you need the personal relationship if you are to go the extra half mile for someone". The strong personal relationship was also felt to help with issues of identity – the members of the group felt that they knew who they were communicating with, even if it was via e-mail. Because they felt that they knew their partners so well, they also had confidence in what they were receiving from them. This point of confidence also has a bearing on what makes a CoP and what differentiates it from a team, for confidence is closely entwined with legitimation. As members get to know each other, have confidence in each other and trust each other, they gain legitimation in the eyes of each other.

One of the most difficult parts of operating in a distributed environment may well be the facilitating of the evolution of the community and the development of the relationships. The case study supported this view as it emphasised the continued importance of maintaining face-to-face contact. The evolution of the CoP was a direct result of face-to-face meetings and the development of the relationships.

**Legitimation**

The group considered themselves a CoP and justified this by giving their definition:

A CoP:
- has a common set of interests motivated to do a common set of something;
- is concerned with motivation;
- is self-generating;
- is self selecting;
- is not necessarily co-located;
- "has a common set of interests motivated to a pattern of work not directed to it."

The main point that comes out of this is that they do not consider CoPs to be (initially at least) formally created. In a formal group such as a team, virtual team or project group, the legitimation of the members comes from the formal structure of the group. In a CoP, the legitimation comes from the social relationships that develop. As people get to know each other, they have confidence in the information and knowledge they receive from their partners. This shows the human aspect of a CoP to be of major importance and therefore does not preclude a formally constituted group or team from developing or evolving into a CoP as the members find they develop relationships, get to know each other well and "go the extra half-mile" beyond the formal relationships. This shows that the essential factor which differentiates a CoP from a team is the "human aspect", that is, the social relationships which are formed in a CoP.

**Peripherality**

In both case studies, there was evidence of a physical peripherality as there were co-located cores with members who were situated elsewhere. In the second case study, there were the two co-located cores with one member in Japan. This member was accepted as a member of the group but did not feature so much in the meetings because of the time difference. She was kept informed of plans and progress but she was not able to play as full a part as other members were. If she wanted to take part in an electronic conference, she had to participate in the middle of the night. She was regarded as being a full member of the group, the other members had every confidence in her ability but the physical and temporal distance meant she was in some ways a peripheral member.

**Participation**

The learning that was seen in the second case study was not explicit. People learnt from each other by collaborating, asking for help, solving problems together. This was perhaps a result of the function of this particular CoP, that is, the members were all experienced specialists in areas of the field and any newcomer would not have to learn a lot of domain knowledge. Rather they would have to learn how the group functioned, the language of the community, how work is done. They would also be able to add to their domain knowledge by collaboration and problem solving. In this case, the face-to-face element is not so essential. The collaboration can be reproduced electronically via shared resources such as documents. Where the face-to-face element did prove advantageous was where problems were solved and help given as a result of informal *ad hoc* encounters. This tended to take place within the cores. *Ad hoc* communication between the cores was a much rarer occurrence.

**Shared documents**

The other major point of interest which came out of the case studies was the use of a shared artefact, in this case a planning document to communicate and share soft knowledge within the community but across national and cultural boundaries. The use of the document acted as a catalyst (as opposed to a vehicle) for the group members to apply their domain and soft knowledge for planning, for reflection, for discussion of issues and for solving problems. The shared document was not essential to their work but it played more roles more importantly than they had previously realised. This particular planning document has undergone a further iteration and still plays a major role in their work. The use of the shared document was interesting in its multiple roles, in particular for its role in the creation and representation of knowledge. However, it raises the question as to what form the knowledge takes, that is, whether it is a "hard" representation of the community members' soft knowledge as their knowledge is embedded in it. If this is the case then it should be a simple matter to go in the other direction and extract the soft knowledge from it, but this does not work. This parallels the artefacts in Hutchins' (1990, 1995) team navigation where he describes them as having the knowledge of previous generations of mariners embedded in them. Star's (1989) boundary objects illustrate that artefacts still need to be interpreted, that is, you still need some domain knowledge. A newcomer would be able to perhaps understand what is meant by an artefact or use it to some degree but an old-timer would be able to get a lot more out of it, fully understand it, use it to the full and perhaps make new inferences using his/her tacit knowledge.

The boundary objects might prove to be an interesting avenue for supporting situated learning in a distributed CoP. In developing the notion of boundary objects, Star (1989) and Star and Griesemer (1989) based them on the notion of immutable mobiles. These are artefacts which do not change but which are able to convey information over a distance. The boundary objects are therefore robust enough to travel between communities but which also have local interpretations. In the second study, we had a different type of boundary – cultural and national boundaries and the boundary between the cores. However there is a shared background within the community and the members all have a high degree of domain knowledge so can get something out of the artefact more than a newcomer could. As boundary objects can be artefacts of all kinds, such as documents or even concepts, Robinson (1997) included procedures. These have the knowledge of previous experts embedded in them. They help newcomers to a community, as the newcomer has to follow the procedure to get the work done. An old-timer on the other hand would have the experience and tacit knowledge to know when to circumvent the procedure, to "break" it or even change it in order to improve it.

Although the shared artefact does not solve the problem of soft knowledge sharing in a distributed international environment the study has shown that it can be of real benefit and can play a variety of useful roles to support the sharing of soft knowledge.

**Conclusion**

CoPs are becoming recognised as being groups within which the sharing of soft knowledge takes place and therefore it is important that they are supported in organisations if KM is to move beyond the established practices of capturing and codifying hard knowledge. The importance of international business means that knowledge now needs to be shared in a distributed environment. The research outlined in this paper has demonstrated that CoPs can function in a distributed environment although the ones found in these case studies were not totally distributed but had co-located cores. It has also shown that a face-to-face element is necessary to take the evolution of the community further more quickly. The case studies have also highlighted an area, the use of shared artefacts, that may be able to contribute to further supporting these communities in their collaboration across time and distance.